\documentclass{aa}                           \usepackage{natbib,twoopt}
\usepackage[breaklinks=true]{hyperref}
\usepackage{amsmath}
\makeatletter
\newcommandtwoopt{\citeads}[3][][]{\href{http://adsabs.harvard.edu/abs/#3}%
  {\def\hyper@linkstart##1##2{}%
    \let\hyper@linkend\@empty\citealp[#1][#2]{#3}}}
\newcommandtwoopt{\citepads}[3][][]{\href{http://adsabs.harvard.edu/abs/#3}%
  {\def\hyper@linkstart##1##2{}%
    \let\hyper@linkend\@empty\citep[#1][#2]{#3}}}
\newcommandtwoopt{\citetads}[3][][]{\href{http://adsabs.harvard.edu/abs/#3}%
  {\def\hyper@linkstart##1##2{}%
    \let\hyper@linkend\@empty\citet[#1][#2]{#3}}}
\newcommandtwoopt{\citeyearads}[3][][]%
                 {\href{http://adsabs.harvard.edu/abs/#3}
                   {\def\hyper@linkstart##1##2{}%
                     \let\hyper@linkend\@empty\citeyear[#1][#2]{#3}}}
                 \makeatother
\usepackage{graphicx}                      \usepackage[hypcap]{caption}
\usepackage{subcaption}         
\usepackage{txfonts}
\begin{document}

   \title{Exploring DCO$^+$ as a tracer of thermal inversion in the disk around the Herbig Ae star HD163296}

   \subtitle{ }

   \author{V.N. Salinas \inst{1}
          \and
          M.R. Hogerheijde \inst{1,2}
          \and
          N.M. Murillo \inst{1}
          \and
          G.S. Mathews \inst{3}
          \and
          C. Qi \inst{5}
          \and
          J.P. Williams \inst{4}
          \and
          D.J. Wilner \inst{5}
          }

   \institute{Leiden Observatory, Leiden University, PO Box 9513,
     2300 RA, Leiden, The Netherlands\\
              \email{salinas@strw.leidenuniv.nl}
    	\and
  Anton Pannekoek Institute for Astronomy, University of Amsterdam, Science Park 904, 1098 XH, Amsterdam, The Netherlands
  	\and
  	University of Hawaii at Manoa, Department of Physics and Astronomy, 2505 Correa Rd., Honolulu HI, USA
  	\and
  Institute for Astronomy, University of Hawaii, 2680 Woodlawn Dr., Honolulu, HI 96826, USA
   \and
    Department of Astronomy, Harvard University, Cambridge, MA 02138, USA
             }

   \date{Received ; accepted }
  \abstract
{In planet-forming disks, deuterated species like DCO$^+$ often
show up in rings. Two chemical formation routes
contribute: cold deuteration at temperatures below 30 K and warm
deuteration at  temperatures up to 80 K.}
{ We aim to reproduce the DCO$^+$ emission in the disk
around HD163296 using a simple 2D chemical model for the formation of DCO$^+$ through
the cold deuteration channel and a parametric treatment of the warm deuteration channel.}
{We use data from ALMA in band 6 to obtain a resolved spectral imaging
data cube of the DCO$^+$ $J$=3--2 line in HD163296 with a synthesized beam of 
0$\farcs$53$\times$ 0$\farcs$42. We adopt a physical structure of the disk 
from the literature that reproduces the spectral energy distribution. We then 
apply a simplified chemical network for the formation of DCO$^+$ that uses the 
physical structure of the disk as parameters along with a
CO abundance profile, a constant HD abundance and a constant ionization rate.  
We model the contribution of the warm deuteration channel with two parameters: 
an effective activation temperature and a constant abundance. Finally, from the 
resulting DCO$^+$ abundances, we calculate the non-LTE emission using the 3D radiative transfer code LIME.
} 
{The observed DCO$^+$ emission is reproduced by a model with cold
deuteration producing abundances up to $1.6\times 10^{-11}$. Warm deuteration, at a
constant abundance of $3.2\times 10^{-12}$, becomes fully effective below 32 K and
tapers off at higher temperatures, reproducing the lack of DCO$^+$ inside 90
AU. Throughout the DCO$^+$ emitting zone a CO abundance of $2\times 10^{-7}$ is found, with $\sim$99\% of it frozen out 
below 19 K. At radii where both cold and warm deuteration are active,
warm deuteration contributes up to 20\% of DCO$^+$, consistent with detailed
chemical models.  The decrease of DCO$^+$ at large radii is attributed to
a temperature inversion at 250 AU, which raises temperatures above
values where cold deuteration operates. Increased 
photodesorption may also limit the radial extent of DCO$^+$. 
The corresponding return of the DCO$^+$ layer to the midplane, together with a radially increasing
ionization fraction, reproduces the local DCO$^+$ emission maximum at $\sim$260 AU.
}
{We can successfully reproduce the observed morphology of DCO$^+$ at large radii by only considering the dependence of temperature in the chemical reactions that produce it. 
Predictions on the location of DCO$^+$ within the disk from simple models depend 
strongly on the gas temperature. Outer disk temperature
inversions, expected when grains decouple from the gas and drift
inward, can lead to secondary maxima in DCO$^+$ emission and a reduction of
its radial extent. This can appear as an outer emission `ring', and
can be used to identify a second CO desorption front.}
   

   \maketitle
%

\section{Introduction}
\label{sec:intro}
DCO$^+$ is a good tracer of the deuterium fractionation and ionization fraction of low temperature environments \citep{Favre2015,Millar1989}. Detections of  DCO$^+$, and other simple deuterated molecules, towards protoplanetary disks are present only in a handful of T Tauri disks: TW Hya \citep{vanDishoeck2003}, DM Tau \citep{Guilloteau2006,Teague2015}, AS 209, IM Lup  V4046 Sgr and LkCa 15  \citep{Oberg2010,Oberg2011,Oberg2015,Huang2017}, and in the disks around the Herbig Ae stars MWC 480 \citep{Huang2017} and HD
163296 \citep{Qi2008,Mathews2013,Qi2015,Yen2016,Salinas2017}. High angular resolution observations of some of these disks have revealed a surprisingly complex radial structure. The chemistry involved in the gas-phase formation of DCO$^+$ is thought to be well understood and has been previously studied in the ISM and in disks \citep{Turner2001,Willacy2007,Roueff2013,Favre2015}. In this paper, we attempt to determine whether our current understanding of the DCO$^+$ chemistry is sufficient to reproduce the complex radial structure seen in protoplanetary disks, particularly in the disk surrounding the Ae Herbig star HD 163296.

DCO$^+$ forms in the gas phase through two different regimes: a low temperature deuteration (henceforth cold deuteration, CD) and a warm deuteration (henceforth  warm deuteration, WD) channel. The CD and WD regimes are gradually enhanced at temperatures lower than $\sim$30 K and $\sim$80 K, respectively \citep{Millar1989,Albertsson2013}. This enhancement is a direct consequence of the lower zero-point vibrational energy of simple deuterated molecules in comparison to their non-deuterated counterparts. The origin of highly deuterated  species, and of DCO$^+$, in the solar nebula can be attributed to in situ synthesis in the primordial disk or to being inherited from the interstellar medium (ISM). Deuterium is injected into the chemistry via ion-molecule reactions, and is kept in it by the endothermic nature of the inverse reaction. For an effective enhancement of the deuterium fractionation ratio the environment must be cold (typically tens of degrees Kelvin), and ionized. In the dense shielded ISM, the ionization environment is mainly dominated by galactic cosmic rays (CRs). 
In the cold and dense outer regions of  protoplanetary disks, where deuterium enrichment takes place, the ionization source comes from galactic cosmic rays (CRs) and to a lesser extent by radiation both from the host star and from external sources in low density regions above the midplane where the ultraviolet (UV) radiation can penetrate. 

DCO$^+$ was proposed first as a good CO snowline tracer in protoplanetary disks \citep{Mathews2013}. Recent chemical models (including the WD channel) conducted by \citet{Favre2015} have proposed DCO$^+$ as a good tracer of the ionization degree in the inner regions of planet-forming disks rather than the temperature structure and the CO snowline. The WD channel formation channel involves ionized simple hydrocarbons, unlike the CD involving H$_2$D$^+$ which is highly reactive with CO. If the CD dominates the formation of DCO$^+$ in disks then it can be used as an indirect tracer of the CO snowline because of its parent molecule, H$_2$D$^+$, is readily destroyed by CO in the gas-phase. The models of \citet{Favre2015} show that the WD channel enhances the column density of DCO$^+$ by a factor of 5 in the warm regions of a T Tauri-like disk, where CO is still in the gas phase, and is responsible for the bulk of the abundance. Observations of significant emission of DCO$^+$ in the inner parts of protoplanetary disks have been already reported \citep{Qi2015,Huang2017}. In particular, \citet{Salinas2017} have seen  DCO$^+$ extending from $\sim$50 AU to $\sim$300 AU in the disk surrounding the Herbig Ae star HD 163296.

HD 163296 has a massive (0.089 M$_\odot$) inclined (44$^o$) disk and its gaseous content, probed by $^{12}$CO emission, extends at least to $\sim$500 AU \citep{Qi2013,Mathews2013}. These attributes and its proximity \citep[122 pc, ][]{Ancker1998} make it an excellent candidate to study both the radial and vertical distribution of deuterated species such as DCO$^+$. The disk has been observed in the millimeter regime revealing a distribution of mm-sized grains that extend up to $\sim$230 AU with multiple rings and gaps \citep{Isella2016}. 

The DCO$^+$ radial distribution of the HD 163296 disk has been characterized by \citet{Salinas2017} using ALMA observations. The data are consistent with three regimes of different constant abundances defined by one inner radius at 50 AU and two breaks at 120 AU and 245 AU. They found that the first two regimes correlate well with the expected WD and CD channels traced by DCN and N$_2$D$^+$. The third regime correlates with the extent of the mm-size dust grains which hints at a local decrease of UV opacity allowing photodesorption of CO and consequent DCO$^+$ formation as proposed for the disk around IM Lup \citep{Oberg2015}. An interesting alternative to explain this excess DCO$^+$ emission is releasing CO through thermal desorption \citep{Cleeves2016}. Dust grain evolution models by \citet{Facchini2017} have predicted a thermal inversion of the dust temperature as a direct consequence of radial drift and settling in the disk surrounding HD 163296 given low turbulence values. 

Our main goal is to implement  a  simple  chemical  network for the CD channel and a parametrized WD channel, to reproduce both the location and the amount of the observed DCO$^+$ in HD163296. Specifically, we aim to constrain the relative contribution of the WD channel after taking into account the CD channel.  Although several other effects can produce the observed ringlike feature at 245 AU and drop off at larger radii of the DCO$^+$ emission, this paper only focuses on the chemical conditions that can lead to such structures. A fully self-consisting modelling including an accurate determination of the temperature profile, dust properties and gas density profile is needed to distinguish between the possible origins of the DCO$^+$ morphology.

In Section  \ref{sec:met} we briefly describe the data and explain our modeling strategy.   Section \ref{sec:res}  contains  the results obtained from our different modeling approaches.  Section \ref{sec:dis} discusses
the  validity of  these models  and provides an interpretation  for the models' parameters.  Finally in section \ref{sec:con} we 
summarize our findings and conclusions.


\section{Methods}
\label{sec:met}
\subsection{Previous observation of HD 163296}

This study uses data of DCO$^+$  $J$=3--2 at 216.112 GHz in the disk surrounding the Herbig Ae star
HD163296    ($\rm\alpha_{2000}$     =    ${\rm    17^h56^m51{\fs}21}$,
$\delta_{2000}$ =  $-21^\circ57'22{\farcs}0$) obtained by the  Atacama Large
(sub-)Millimeter Array (ALMA)  in Band 6 as  a part of Cycle  2 on  2014 July
27-29 (project  2013.1.01268.S). The spectral resolution is 0.7 MHz corresponding to 0.085 km s$^{-1}$ with respect to the rest frequency of the line. The $uv$-coverage of the data ranges from 20 to 630 k$\lambda$.

\begin{figure}[!h]
        \centering
        \includegraphics[width=0.98\columnwidth]{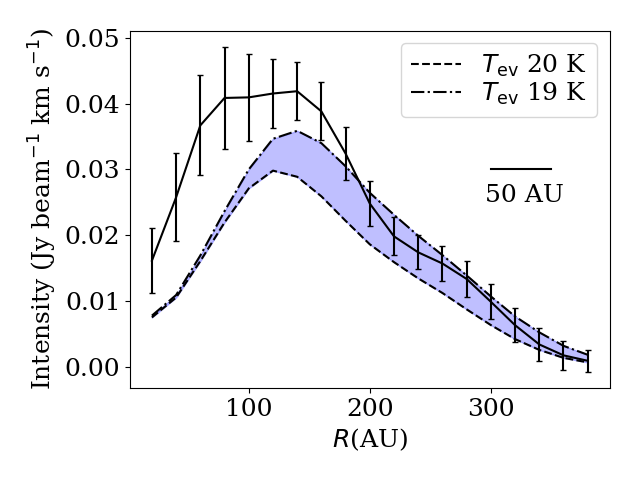}
        \caption{The black continuous line shows the observed DCO$^+$ radial profile. These profiles are obtained by averaging the values of concentric ellipsoids of the integrated intensity maps. The error bars of the observed DCO$^+$ radial profile corresponds to 3$\sigma$, where $\sigma$  is the standard deviation of the values contained in the ellipsoid divided by  the square root of the number of beams. The 50 AU bar corresponds to the semi-minor axis of the synthesized beam and serves as a measure of the spatial resolution. The dashed lines show radial profiles of the emission from models with two different evaporation temperature of CO $T_{\rm CO}$=19-20 K and a constant abundance $X_{\rm in}=$5.0$\times10^{-5}$}
\label{fig:mod:cold}
\end{figure}

 The data were reduced as described by \citet{Salinas2017} and \citet{Carney2017}.   The  DCO$^+$ $J$=3--2 line was continuum subtracted in the visibility plane using a first-order polynomial fit and imaged using a Briggs weighting of 0.5. The resulting synthesized beam has dimensions of 0$\farcs$53$\times$0$\farcs$42 and is shown along with the obtained channel maps in Appendix \ref{appendix}. \citet{Salinas2017} use a Keplerian mask to obtain an integrated intensity map. A radial emission profile can be constructed by taking the average value of concentric ellipsoids, centered at the star, in the integrated intensity map. The resulting DCO$^+$ radial emission profile can be seen in Fig.~\ref{fig:mod:cold}. 
The observational analysis of \citet{Salinas2017} modeled DCO$^+$ as a three-region radial structure. They argue that these radial regions correspond roughly to the WD channel, the CD channel and a third unidentified regime at larger radii ($\gtrsim$240 AU). Their analysis treated each region independently and assumed that the bulk of the DCO$^+$ emission comes from the midplane. These assumptions do not impact their ability to constrain the regimes radially, but do not allow to constrain the contribution of the WD channel in the HD163296 disk. Whereas \citet{Salinas2017} only derived vertically averaged DCO$^+$ abundances, here we carry out fully 2D (radius and height) modeling of the chemistry. Because deuteration, and the formation of DCO$^+$, is a temperature dependent process, a vertical treatment of its abundance is needed in disks due to their strong vertical temperature gradient. For our analysis, we will assume that the WD channel remains active where the CD channel operates, but its contribution to the DCO$^+$ emission in the outer regions of the disk is negligible. If this is the case, reproducing the emission at large radii alone can provide a constraint on the amount of DCO$^+$ produced by the CD channel.

\subsection{Chemical model}
\label{sec:chem}
Deuterium is incorporated into the gas chemistry mainly through the following
ion molecule reactions \citep{Gerner2015,Turner2001}
\begin{subequations}
\begin{equation} \rm H_3^+ +HD\leftrightharpoons H_2D^++H_2+230K{\rm , }
\label{eq:low:T}
\end{equation}
\begin{equation} \rm CH_3^+ +HD\leftrightharpoons CH_2D^+ +H_2+370K{\rm , }
\label{eq:high:T}
\end{equation}
\begin{equation} \rm C_2H_2^+ +HD\leftrightharpoons C_2HD^+ +H_2+550K{\rm . }
\label{eq:high:T:2}
\end{equation}
\end{subequations}
The right-to-left reactions  of Equations~\ref{eq:low:T}, \ref{eq:high:T} and \ref{eq:high:T:2} are endothermic
and effectively enhance deuterium fractionation in low temperature environments.  Equation ~\ref{eq:low:T}  corresponds  to  the so-called  CD channel and are active at  temperatures   ranging    from   10-30    K \citep{Millar1989,Albertsson2013}. Eq.~\ref{eq:high:T} and \ref{eq:high:T:2},
involving  light  hydrocarbons, correspond to the WD channel 
and is active at warmer temperatures ranging from 10-80 K.

DCO$^+$ is formed in the gas-phase involving both the CD and WD channels through the reactions \citep{Watson1976,Wootten1987,Favre2015}
\begin{subequations} 
\begin{equation}\rm H_2D^++CO\longrightarrow DCO^++H_2\rm{,}
\label{eq:DCOp_1}
\end{equation} 
\begin{equation} \rm CH_2D^++O\longrightarrow DCO^++H_2\rm{,}
\label{eq:DCOp_2}
\end{equation}
\end{subequations} 
or with products of CH$_2$D$^+$ such as CH$_4$D$^+$ 
\begin{equation} \rm CH_4D^++CO\longrightarrow DCO^++CH_4\rm{.}
\label{eq:DCOp_3}
\end{equation}

We will model the CD channel using a simple chemical network, in 2D, and regard the WD channel as a constant abundance ($X_{\rm WD}$) contribution that occurs at temperatures lower than an effective  temperature ($T_{\rm eff}$). The DCO$^+$ chemical network involving the CD channel can be boiled down to only ten chemical reactions. This system can be solved analytically as proposed by \citet{Murillo2015}. We use their prescription and simplified chemical network of the CD channel and apply it to HD163296. The input parameters are: the gas density ($n$(H$_2$)), the gas temperature ($T_{\rm gas}$), the CO gas abundance ($X$(CO)) and the HD gas abundance ($X$(HD)). The ionization rate ($\zeta$) is constant throughout the disk and equal to$ 1.3\times10^{-17}$ s$^{-1}$. The ortho to para ratio of H$_2$ is considered to be in thermal equilibrium (LTE) and is approximated by the following expression,
\begin{equation}
\frac{\rm o}{\rm p}=9\hspace{0.1cm}{\rm exp}\left(- \frac{\rm 170 K}{T}\right).
\end{equation}
We use a lower limit of 10$^{-3}$ at low temperatures as  described in \citet{Murillo2015}. They contrasted their results to a full chemical network including gas-grain balance (freeze-out, thermal desorption, and cosmic-ray-induced photodesorption) confirming the general trend found by the simplified network. The advantage of using this simple network over a full chemical calculation, is that it allows us to more easily investigate the dependency of the DCO$^+$ emission on individual model characteristics.   Detailed chemical modeling, including the reactions shown in Eq.~\ref{eq:high:T} and \ref{eq:high:T:2} would be needed to further investigate the DCO$^+$ radial distribution, but this is beyond the scope of this paper.

\subsection{Implementation}

\begin{figure*}[]
\centering
\begin{subfigure}{0.49\textwidth}
        \centering
        \includegraphics[width=0.98\textwidth]{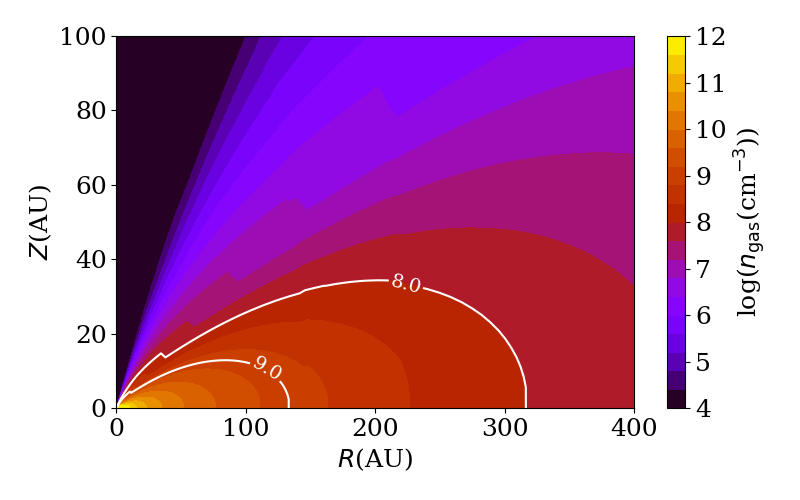}
\end{subfigure}
\begin{subfigure}{0.49\textwidth}
        \centering
        \includegraphics[width=0.98\textwidth]{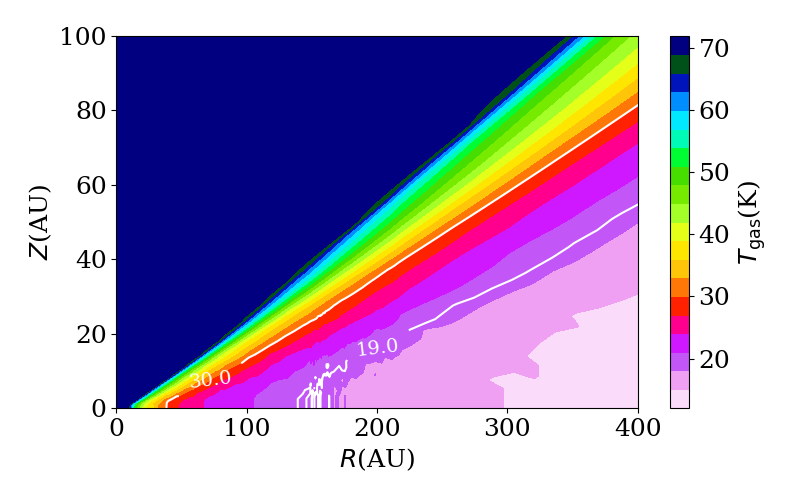}
\end{subfigure}
\caption{Visualization of the adopted physical model. The left panel shows the model density profile with white contours at 10$^8$ cm$^{-3}$ and 10$^9$ cm$^{-3}$. The right panel shows the gas temperature structured clipped at 80 K to enhance the color gradient in the region of interest and white contours at 20, 30 and 80 K.}
\label{fig:phy:mod}
\end{figure*}

We adopt the gas density and dust temperature from the physical model used by
 \citet{Mathews2013} as inputs for the simple chemical network describe above. This parametric model is an approximation of the model used
 by \citet{Qi2011} which fits both the SED and their milliliter observations. The density structure is defined by
\begin{equation*}
\label{eq:sigma}
\Sigma_d(R)=
\begin{cases}
\Sigma_C~\left(\frac{R}{R_c}\right)^{-1}{\rm
  exp}\left[-\left(\frac{R}{R_c}\right)\right] & \text{if }R \geq R_{\rm in} \\
0 & \text{if }R < R_{\rm in ,}
\end{cases}
\end{equation*} 
where $\Sigma_{\rm C}$ is  determined by the total disk  mass $M_{\rm disk}$
(0.089 M$_{\odot}$), $R_{\rm C}$ (150  AU) is the characteristic radius
and $R_{\rm in}$ (0.6  AU) is the inner rim of  the disk. The vertical
structure is treated as a  Gaussian distribution with an angular scale
height defined by
\begin{equation*}
h(R)=h_{\rm C} \left(\frac{R}{R_{\rm C}}\right)^\psi ,
\end{equation*} 
where $\psi$ (0.066) is the flaring  power of the disk and $h_{\rm C}$
is  the angular  scale  height at  the  characteristic radius  $R_{\rm
C}$ that can take different values for the gas and the dust distribution \citep[see appendix  A;][]{Mathews2013}.  The dust  temperature
profile  was  computed  by  the   2D  radiative  transfer  code  RADMC
\citep{Dullemond2004b} and is shown in Fig.~\ref{fig:phy:mod} along with the gas density profile.

In addition to the gas density and dust temperature structure, CO and HD abundance profiles are required as inputs for the simple chemical network. We use a CO abundance profile described by three parameters:  an effective dust temperature where CO starts to evaporate and becomes optimal for DCO$^+$ formation ($T_{\rm CO}$), and two constant abundances for gas-phase CO inside ($X_{\rm low}$) and outside ($X_{\rm high}$) the freeze-out zone. We set $X_{\rm low}$=0.01$X_{\rm high}$ for all our models. We assume that the dust temperature ($T_{\rm dust}$) equals the gas temperature ($T_{\rm gas}$), which is a reasonable assumption in the dense regions that we  focus on. We set the HD abundance, with respect to the nuclei density $n_{\rm H}$=2$n_{\rm H_2}$, constant through the entire disk and equal to the cosmic D/H ratio $\sim$10$^{-5}$ \citep{Vidal-Madjar1991}. 

\subsection{Radiative transfer}
We used LIME (v1.5), a 3D  radiative transfer code in non-LTE \citep{Brinch2010} to compare our DCO$^+$ observations to the DCO$^+$ abundance model obtained using the prescription described above. LIME can
produce line and continuum radiation from a physical model, an abundance distribution and the rate coefficients for a given molecular transition.  We
use  the  rate  coefficients  from the  Leiden  Atomic  and  Molecular
Database
\citep{LAMBDA_database2005}\footnote{www.strw.leidenuniv.nl/\textasciitilde moldata/} for  DCO$^+$. These are 
the same  collision rates as those listed for HCO$^+$ \citep{Flower1999}. The Einstein A
coefficients taken are from  CDMS and JPL. We use a 
grid of 100000 points that are created applying a weighted random selection in $R$ using a logarithmic
scale. This weighted random selection favors denser gas regions and will always select a random point where the DCO$^+$ abundance of the model is higher than 10$^{-15}$.  Establishing  a convergence  criteria on encompassing all  of the
grid points is difficult. We manually  set the number of iterations to
20 and  confirm convergence  by comparing consecutive  iterations.

The resulting model cube is continuum subtracted using the first channel as a continuum estimator. Then, each spectral plane of the model is convolved with the synthesized beam of the DCO$^+$ data cube. This is equivalent to simulate the visibilities from a sky model since the $uv$-space is well sampled (see the result in Appendix~\ref{appendix}).

\section{Results}
\label{sec:res}
\subsection{Standard model (CD)}	
\label{sec:CD}
 We chose a standard CO abundance model with a constant abundance $X_{\rm high}$=5.0$\times10^{-5}$ at temperatures above an evaporation temperature of  $T_{\rm CO}$=19 K. This evaporation temperature corresponds to $\sim$150 AU at the midplane in our adopted temperature structure and correlates well with the second emission ring thought to be produced by the CD channel \citep{Salinas2017}. The adopted CO abundance is similar to previous models of CO isotopologues \citep{Qi2015,Carney2017}. We also include a radial cut off at 300 AU where the DCO$^+$ abundance drops to zero because the emission disappears at $\sim$300 AU. We  further discuss this value  in Sec.~\ref{sec:dis:out}.

The resulting DCO$^+$ $J$=3--2 emission radial profile for the standard model is shown in Fig.~\ref{fig:mod:cold}. The figure also shows our standard model with $T_{\rm CO}$=20 K keeping $X_{\rm high}$ at the same value to illustrate the model dependency on this parameter. Note that by changing the CO abundance the model is capable of compensating the difference in evaporation temperature because these two parameters are degenerate. A higher CO abundance will produce less DCO$^+$ and vice versa. The adopted CO abundance of $5\times10^{-5}$ correspond to a moderate carbon depletion consistent with other warmer disks such as HD100546 \citep{Kama2016}.

We can obtain an estimate of the emission contribution from the WD channel subtracting the standard model from the data. We perform a channel by channel subtraction from the data cube and the convolved model cube to calculate a residual cube (see example in Appendix \ref{appendix}).  Figure~\ref{fig:mod:abu:cold} shows the residual radial curve from the standard models with $T_{\rm CO}=$19 K and $T_{\rm CO}=20$ K and their correspondent abundance estimate. The residual radial profile is obtained applying a Keplerian mask \citep[see Appendix in][]{Salinas2017} to the residual cube and each of the radial bins correspond to concentric ellipsoids projected to be equidistant to the central star.  At $R\gtrsim$ 180 AU we can  only provide an upper limit of about a few 10$^{-12}$. At $R\lesssim$ 180 AU the abundance is a few 10$^{-13}$ and starts declining at $\sim$50 AU. 

The radial abundance estimate is calculated assuming LTE and that the emission is optically thin. If we regard the emission as coming from an isothermal medium in LTE we can calculate the correspondent column density at different radii assuming an excitation temperature by using the analytical formula of \citet{Remijan2003}. This 1D analysis was previously performed by \citet{Salinas2017} and we use the same prescription and excitation temperature profile for DCO$^+$. Finally, we divide the column density estimate by the surface density profile in Eq.~\ref{eq:sigma} to get a vertically averaged abundance. The radial abundance profile only provides a lower limit to the actual DCO$^+$ because it emits from a layer  set by the activation temperature of the CD and WD channel and the CO evaporation temperature constraining its vertical extent. The simple chemical model of the CD channel can reproduce the DCO$^+$ emission in the disk around HD163296  at large radii ($R>180$ AU), but requires additional DCO$^+$, of a few 10$^{-12}$ in abundance, to account for the inner emission produced by the WD channel and a radial cut-off at $\sim$300 AU.

\begin{figure}
\begin{subfigure}{0.98\columnwidth}
        \centering
        \includegraphics[width=0.98\columnwidth]{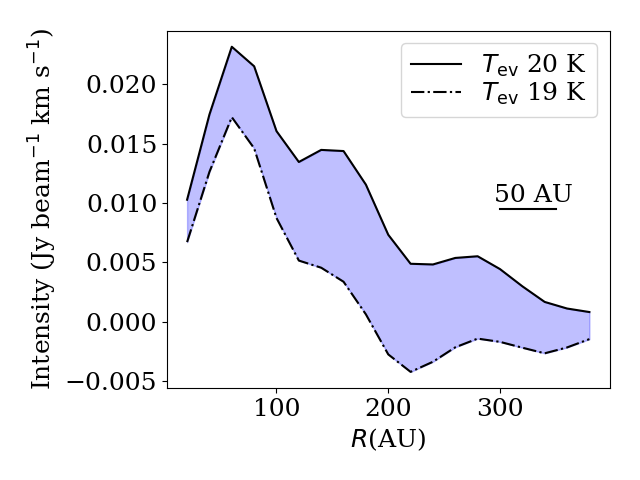}
\end{subfigure} 
\begin{subfigure}{0.98\columnwidth}
        \centering
        \includegraphics[width=0.98\columnwidth]{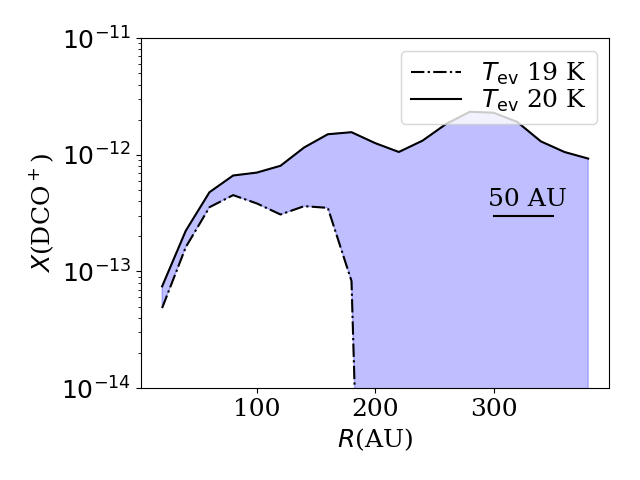}
\end{subfigure} 
\caption{The top panel shows the residual radial profiles from the models shown in Fig.~\ref{fig:mod:cold} with an evaporation temperature of CO $T_{\rm CO}$=19K-20K and a constant abundance $X_{\rm in}=$5.0$\times10^{-5}$. The bottom panel shows a vertically averaged abundance estimate of the radial curves of the top panel using the same excitation temperature profile and prescription proposed by \citet{Salinas2017} to convert emission to column densities. The 50 AU bar corresponds to the semi-minor axis of the synthesized beam and serves as a measure of the spatial resolution.}
\label{fig:mod:abu:cold}
\end{figure}

\subsection{Thermal inversion model (CD+TI)}
\label{sec:res:TI}
The standard model uses a radial cutoff to suppress the emission of DCO$^+$ in the outer disk, but a thermal inversion (TI) could prevent the CD channel from being active \citep{Cleeves2016}. Temperatures higher than the activation barrier for the CD (and WD) channel would prevent the reactions shown in Eq. \ref{eq:low:T} and \ref{eq:high:T}, reducing DCO$^+$ formation. A thermal inversion is produced by the dust evolution modeling of \citet{Facchini2017} as a direct consequence of grain growth and settling. Considering both processes  results in a radial decrease of the scale height of the dust because the turbulence is less capable of stirring up the grains as the density drops. This leads to a a radial temperature drop as less stellar light is intercepted. At larger radii, where almost all of the big grains have migrated inward, small dust is stirred to the disk surface, out of the shadow cast in intermediate radii, intercepting more radiation and raising the temperature.

 We modify the temperature structure of our standard model to mimic this effect using the following parametrization
\begin{equation}
\label{eq:TI}
T\sp{\prime}(R)=25K \left(1-{\rm exp}\left(-\frac{R-R_1}{R_2-R_1}\right)^2\right)+T(R),
\end{equation}
where $R_1=240$ AU is the inflection point at which the TI occurs and $R_2=300$ AU a characteristic radius. We choose these values so that the temperature at $\sim$290 AU reaches $>$30 K and effectively blocks the CD channel. We also use a CO abundance parameter of $X_{\rm high}$=2.0$\times$10$^{-7}$, keeping $T_{\rm CO}$ at 19 K, to better match the shape of DCO$^+$ at larger radii. This value is much lower than the 5$\times10^{-5}$ of the fiducial model.

 The adopted simple chemical network reaches a maximum in DCO$^+$ production for CO abundances of $1\times10^{-5}$. Higher values result in lower DCO$^+$ abundances because the gas-phase CO abundance surpasses the assumed HD abundance blocking its reaction with H$_3^+$ (Eq.~\ref{eq:low:T}). Similarly, lower values than $1\times10^{-5}$ result in lower DCO$^+$ abundances because CO in the gas-phase is required for Eq.~\ref{eq:DCOp_1} to proceed. If the gradient of the CO abundance is steep and smaller or comparable to our resolution, we can think of the $X_{\rm high}$ parameter of our simple step abundance model as an effective CO abundance that accounts for the production gradient of DCO$^+$. The thermal inversion parametrization region in our model has a steep temperature gradient that results in a steep second desorption front of CO in the midplane, at $\sim$ 240-300 AU, comparable in size to our resolution. Our preferred value of $X_{\rm high}$=2.0$\times$10$^{-7}$ is an effective CO abundance that reproduces the DCO$^+$ emission for such a steep gradient. 

The first CO desorption front occurs much further in. Our CD+TI model places this desorption front at about $\sim$150 AU, corresponding to a temperature of $\sim$ 19 K in our adopted model, where the CO starts to evaporate with modest abundances of $\sim$2$\times10^{-7}$. Observations of C$^{18}$O and N$_2$H$^+$ place the CO snowline at $\sim$ 90 AU \citep{Qi2015}, corresponding to a temperature of $\sim$ 22 K in our adopted model, where the bulk of CO is released into the gas-phase with abundances of $\sim$5$\times10^{-5}$. Coincidentally, CO abundances of  $\sim$5$\times10^{-5}$ and  $\sim$2$\times10^{-7}$ produce the same amount of DCO$^+$ at temperatures where the CD channel operates. This explains why in Sec.\ref{sec:CD} the CD model reproduces the DCO$^+$ emission with a CO abundance of $\sim$5$\times10^{-5}$ and a desorption front at $\sim$ 150 AU. We find that to reconcile the CO snowline location at 90 AU, with an abundance of 5$\times10^{-5}$, and the observed DCO$^+$ emission our CO abundance profile requires a zone between 90 AU and 150 AU with an effective abundance of 2$\times10^{-7}$ which we interpret as the onset of the thermal release of CO into the gas-phase. 
\begin{figure*}
\begin{subfigure}{0.33\textwidth}
        \centering
        \caption{Model CD}
        \includegraphics[width=0.98\columnwidth]{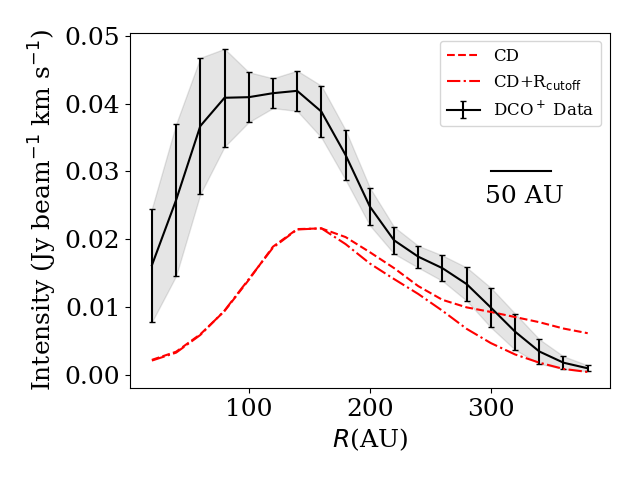}
\end{subfigure}        
\begin{subfigure}{0.33\textwidth}
        \centering
        \caption{Model CD+TI}
        \includegraphics[width=0.98\columnwidth]{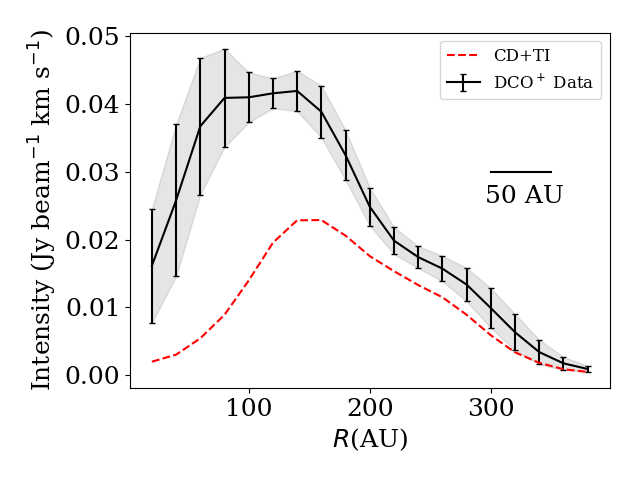}
\end{subfigure} 
\begin{subfigure}{0.33\textwidth}
        \centering
        \caption{Model CD+WD+TI}
        \includegraphics[width=0.98\columnwidth]{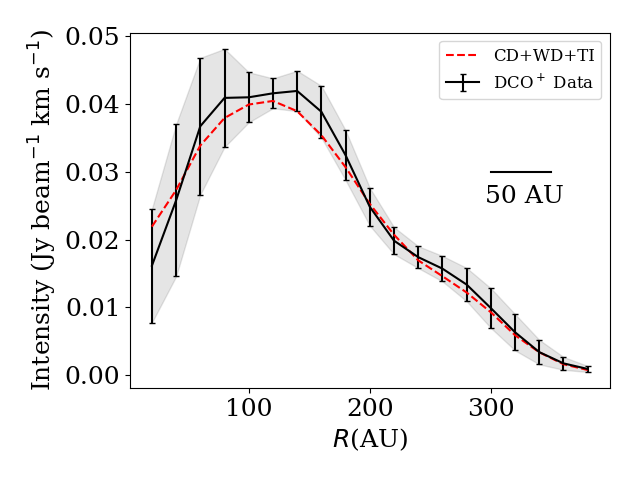}
\end{subfigure} 
\begin{subfigure}{0.33\textwidth}
  \centering
        \includegraphics[width=0.98\columnwidth]{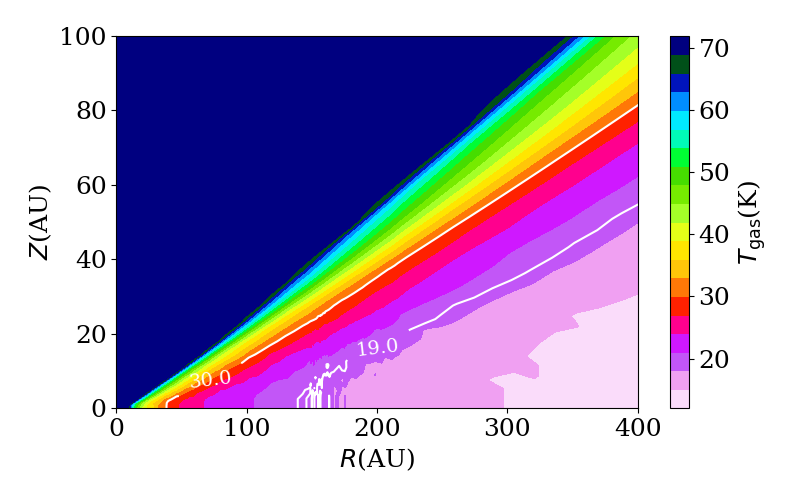}
\end{subfigure}        
\begin{subfigure}{0.33\textwidth}
        \centering
        \includegraphics[width=0.98\columnwidth]{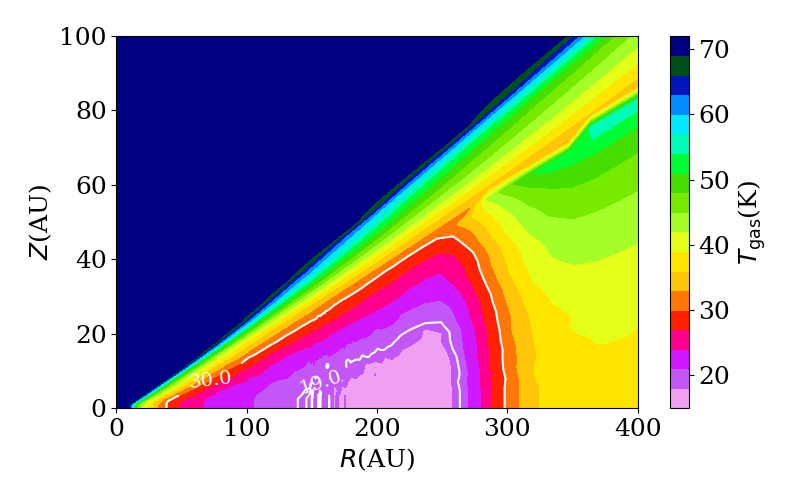}
\end{subfigure} 
\begin{subfigure}{0.33\textwidth}
        \centering
        \includegraphics[width=0.98\columnwidth]{CD+TI_temp.png}
\end{subfigure} 
\begin{subfigure}{0.33\textwidth}
  \centering
        \includegraphics[width=0.98\columnwidth]{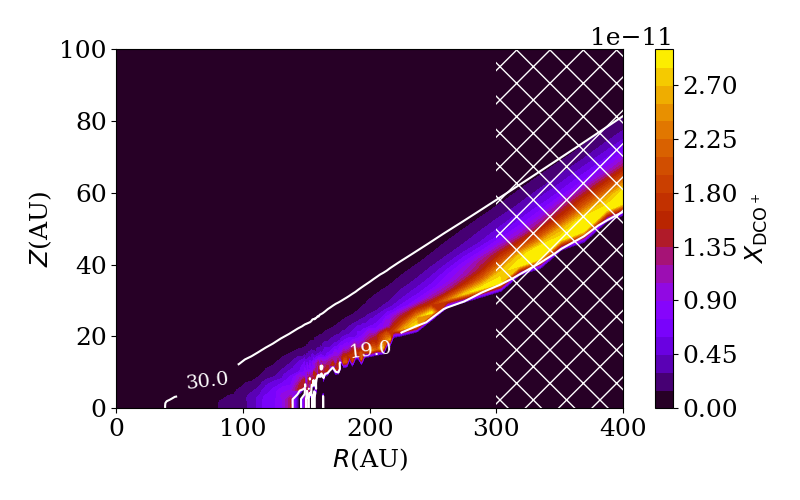}
\end{subfigure}        
\begin{subfigure}{0.33\textwidth}
        \centering
        \includegraphics[width=0.98\columnwidth]{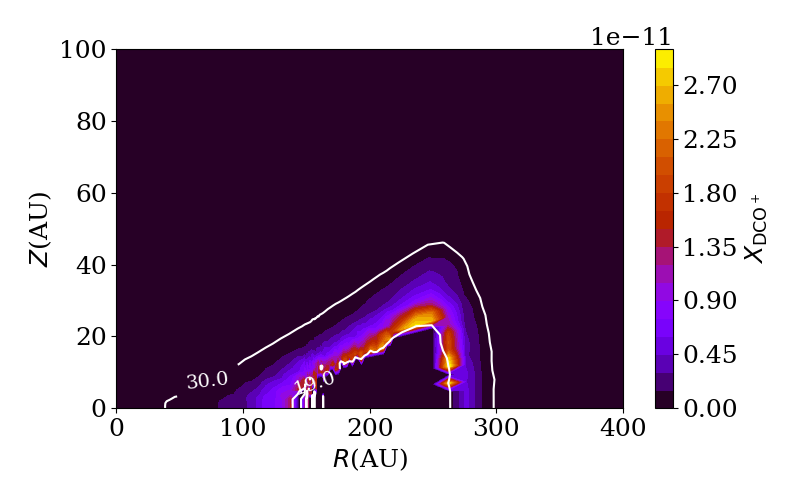}
\end{subfigure} 
\begin{subfigure}{0.33\textwidth}
        \centering
        \includegraphics[width=0.98\columnwidth]{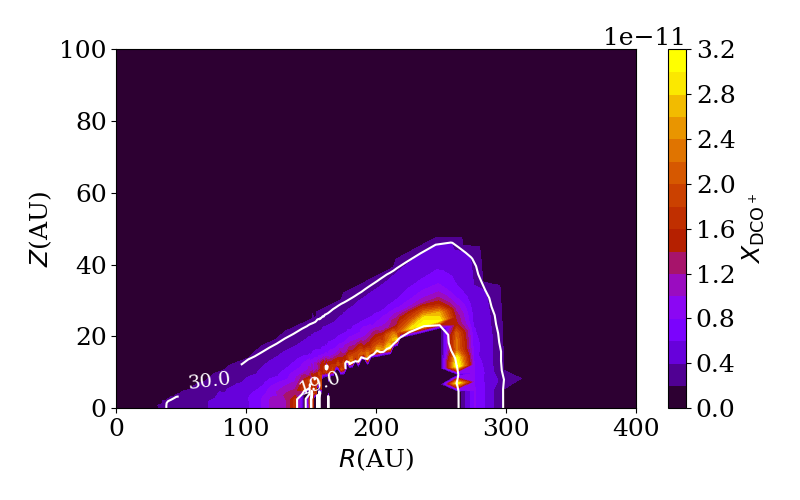}
\end{subfigure}        
\caption{The top panels show the resulting radial profile of the DCO$^+$ integrated intensity map of both models and data.  The error bars on the data corresponds to 3$\sigma$, where $\sigma$ is calculated as the standard deviation of a single ellipsoid divided by the square root of the number of beams. The middle panels show the models' temperature profile with contours at 19 and 30 K. The bottom panel shows the DCO$^+$ abundance profile over-plotted with temperature contours at 19 and 30 K. The hatched region in the DCO$^+$ abundance from model CD marks the radial cut off used in our standard model to reproduce the absence of emission at $R>$300 AU. }
\label{fig:mod:CD+WD+TI}
\end{figure*}

The resulting DCO$^+$ abundance, gas density and emission profile are shown in the middle panels of Fig.~\ref{fig:mod:CD+WD+TI}. DCO$^+$ is confined between 30 K and 19 K, corresponding to the temperature where the CD channel starts to be efficient and to the temperature where CO freeze out passes 99\% respectively. The DCO$^+$ abundance increases radially as a consequence of the radial increase of the ionization fraction which is $\propto \frac{1}{\sqrt{n}}$ if the ionization rate $\zeta$ is constant. The thermal inversion parametrization of this model, coupled with the simple chemical network for CD channel, can reproduce the radial shape of the DCO$^+$ in the disk around HD163296 at $R>180$ AU and the lack of emission at $R>300$ AU.

\subsection{The CD+WD+TI model}

Finally, we add a WD channel component to our CD+TI model by modifying the obtained DCO$^+$ abundance in the following parametrization
\begin{equation}
X_{\rm DCO^+}(r,t)= 
\left\{
	\begin{array}{ll}
		X_{\rm CD+TI}(r,t)+ X_{\rm WD}  & \mbox{if } T(r,t) \leq T_{\rm eff} \\
		X_{\rm CD+TI}(r,t) & \mbox{if } T(r,t) > T_{\rm eff}
	\end{array}
	\right.
\end{equation} 
where $X_{\rm WD}$ is a constant abundance parameter describing the WD channel contribution and $T_{\rm eff}$ the effective temperature for the WD channel at which the the WD channel switches on. The $T_{\rm eff}$ is not the temperature where WD starts to operate (which is 80 K), but rather where reaches its full effectiveness. Just like the CD channel, the WD channel does not switch on abruptly, but gradually increases toward lower temperatures. The right column of Fig.~\ref{fig:mod:CD+WD+TI} show the modified DCO$^+$ abundance, temperature and emission profile of the CD+WD+TI model using $X_{\rm WD}=3.2\times10^{-12}$ and $T_{\rm eff}$= 32 K and the parameters from the CD+TI model. These values are taken to match the inner DCO$^+$ emission at R$\lesssim$ 150 AU. Note that the parameter $T_{\rm eff}$= 32 K is tracing the location of an effective isothermal surface where WD starts to operate. This location is effectively constrained from the data and corresponds to $\sim$ 40 AU in the midplane. The exact value of the parameter $T_{\rm eff}$ cannot be constrained without a gas temperature model and only represents the location of the isothermal surface. On the other hand, the $X_{\rm WD}$ parameter is much better constrained by our modelling because it does not depend strongly on the assumed gas temperature.  The CD+WD+TI model reproduces  simultaneously the inner and outer features of the DCO$^+$ radial profile. The residual (and model) channel maps can be seen in Appendix~\ref{appendix}. 

For comparison, the left column of Fig.~\ref{fig:mod:CD+WD+TI} shows the DCO$^+$ abundance, temperature and emission profile of model CD. The emission profile of model CD does not include the radial cut off of our standard model. This radial cut off is shown as a hatched region in the DCO$^+$ abundance profile of model CD. Without reducing the amount DCO$^+$ at larger radii the CD channel alone overproduces the observed emission. The CD+WD+TI model effectively reproduces the DCO$^+$ radial emission profile, at large and small radii, of the disk around HD163296 including the lack of emission at $R>300$ AU.

\section{Discussion}
\label{sec:dis}

\subsection{DCO$^+$ Outer radius}
\label{sec:dis:out}
Our standard model uses a radial cutoff in the DCO$^+$ abundance to reproduce the absence in DCO$^+$ emission at $R\gtrsim$300 AU. In this section we discuss several possibilities to explain such a drop, namely: a drop in total gas density, a local increase in the o/p ratio of H$_2$, photodesorption or thermal desorption of CO and a higher electron fraction.

First, the emission drop could be the consequence of a drop in total gas density. Observations of CO isotoplogues at very high angular resolution \citep{Isella2016} show that the $^{12}$CO, $^{13}$CO and C$^{18}$O emission abruptly diminish at different radii; $\sim$630 AU, $\sim$510 AU and $\sim$ 360 AU respectively. If C$^{18}$O (the least abundant CO isotopologue) is tracing the total gas density, the radial cutoff of DCO$^+$ might be due to a lack of total gas density at a similar radius at which the C$^{18}$O emission drops. However, isotope selective dissociation of C$^{18}$O is a more likely explanation of its drop off at 360 AU \citep{Miotello2014,Visser2009}. If we adopt a plausible model of the CO column density profile, such as the one obtained by \citet{Facchini2017} (left panel of their Fig.8), the column density of C$^{18}$O at $R\gtrsim$360 AU is not high enough to self-shield from UV radiation assuming the canonical ISM C$^{18}$O/$^{12}$CO ratio of 550. This would be consistent with the different radii at which the emission of $^{12}$CO, $^{13}$CO and C$^{18}$O drop. It is therefore unlikely that the drop off of DCO$^+$ is due to an overall drop in gas surface density. Nevertheless, planet-disk interactions, as probed by the high resolution observations of HD 163296 in the millimiter continuum and CO isotopologues \citep{Isella2016}, can produce features in the dust and gas density profile with spatial scales that are similar to the ones seen in the DCO$^+$ ring-like structures.

A second explanation comes from the difference in the zero-point energies of ortho and para H$_2$. Since o-H$_2$ has more energy than p-H$_2$, the endothermic reaction that introduces deuterium via the latter has a lower energy barrier than the former, enhancing the deuterium fractionation more efficiently at cold temperatures for a thermal o/p ratio of H$_2$. This means that a local increase in the o/p ratio of H$_2$ could result in a drop of DCO$^+$ abundance. However, it is unclear why H$_2$ would have a spin temperature much higher (80 K) than the region where DCO$^+$ forms ($<$30 K).

Thirdly, desorption of CO at large radii, either thermally or through photodesorption, can raise the CO abundance above the value where the formation of DCO$^+$ is quenched. The distinctive feature at $\sim$260 AU, where the DCO$^+$ emission shows a 'bump', is a natural consequence of this. Both in a model with a thermal inversion and in a model with increased photodesorption through increased UV penetration, the layer of DCO$^+$ folds back to the midplane. Increased excitation and column density lead to a maximum in emission. The presence of this 'bump', in our data, therefore strongly supports this scenario. Small scale structures in CO abundance, gas temperature or H$_2$ o/p ratio could also produce the observed emission excess at $~$250 AU, but not as naturally as a thermal inversion. 
An extreme case of this scenario can also explain the DCO$^+$ emission rings seen in other disks such as IM Lup \citep{Oberg2015}. If the inflection point of the temperature profile occurs at large heights, at low gas densities, the emission could hide below the noise level and come back again at larger radii where the DCO$^+$ comes back to the midplane. This would give the illusion of two DCO$^+$ rings at low sensitivities.

The shape of our modified temperature profile resembles qualitatively the models of \citet{Facchini2017} but their results are heavily model dependent, in particular to the turbulent parameter $\alpha$. The thermal inversion effect is most pronounced at lower alpha parameters (10$^{-3}$-10$^{-4}$). The temperature inversion in these models is smoother than our proposed parametrization. The CO ice in their models is confined from $\sim$200 to $\sim$400 AU, with $\alpha=$10$^{-4}$, whereas our models confine CO ice from $\sim$150 AU to $\sim$260 AU. A different temperature structure with a slightly hotter disk could also shape the CO ice region differently. Our constraining CD+WD+TI model applied to the DCO$^+$ data reveals the location of the thermal inversion and the temperature structure at large radii.

Instead of thermal desorption, photodesorption of CO by increased UV penetration can yield a similar cutoff to the DCO$^+$ emission profile. This has been invoked for the DCO$^+$ ring seen in IM Lup and  full chemical models, of typical disks around T Tauri stars, show that this is a plausible scenario \citep{Oberg2015}. Our modeling cannot distinguish this from thermal desorption. Additional observations are needed, either constraining the temperature (e.g via multiple transitions of the rotational lines of optically thin species such as H$_2$CO \citep{Carney2017}), or through UV tracers and/or modeling of the dust. Both thermal and photodesorption could be occurring, potentially creating even more complex structures.

 Finally, a higher electron density at radii $R>300$ AU cannot significantly destroy DCO$^+$ resulting in the observed absence of emission. Our DCO$^+$ models cannot constrain the electron density directly because it is degenerate with gas-phase CO abundance. However, if no thermal inversion is invoked and CO remains in the gas-phase above 19 K (model CD), the required ionization rate to completely quench the DCO$^+$ production at $R>300$ AU is at least 8 orders of magnitude higher than the assumed canonical value of 1.36$\times 10^{-17}$ s$^{-1}$. This implies an electron abundance of a few 10$^{-4}$, much higher than expected for the warm molecular layer. 

\subsection{Low vs high temperature deuteration pathways}
\label{sec:dis:warm}
 
Our preferred CD+WD+TI model uses a constant abundance $X_{\rm WD}=3.2\times10^{-12}$ and an effective temperature of $T_{\rm eff}$= 32 K to parametrize the contribution and onset of the WD channel. From Fig.~\ref{fig:mod:CD+WD+TI} the abundance of model CD is about $1.2\times 10^{-11}$ at larger radii corresponding to a ratio between the DCO$^+$ column densities produced by the WD and the CD channel of ~0.2. This ratio agrees well with the detailed chemical modeling of \citet{Favre2015} which is on average about 0.25 outside the CO snowline. Since the amount of DCO$^+$ produced by the CD channel depends on the CO abundance, an appropriate CO gradient can maximize the DCO$^+$ produced via the CD channel and minimize the contribution of the WD channel to $\sim$ 10\%. This limit is only slightly lower than the 20\% found by our CD+WD+TI model, which is therefore a robust number. The innermost drop off of the DCO$^+$ emission is probably caused by the dust becoming optically thick in the inner 50 AU \cite{Isella2016}, and not by the WD channel becoming fully operative. This means that our $T_{\rm eff}$ is not tracing the onset of the WD channel but only the radius at where the dust becomes optically thick. On the other hand, our $X_{\rm WD}$ parameter is independent from this effect. Provided that the contribution of the WD channel can be constrained, DCO$^+$ is a promising tracer of  the temperature structure and CO snowline of a protoplanetary disk through their CD channel formation pathway.


\section{Summary}
\label{sec:con}

In this work, we implemented a simple chemical network for cold deuteration and a parametrized treatment of warm deuteration, and carry out a 2D modeling of the DCO$^+$ emission in the disk around HD163296. The following points summarize the conclusions of this work:
\begin{itemize}

\item We found that a simple chemical model of the CD channel using a CO constant abundance of 2.0$\times$10$^{-7}$, above an effective dust temperature of 19 K where CO starts to evaporate, coupled with thermal inversion at around $\sim$260 AU can reproduce the DCO$^+$ emission in the outer regions of the disk surrounding HD163296.

\item In addition, modeling the contribution of the WD channel with constant abundance of 3.2$\times 10^{-12}$ and an effective temperature of 32 K, describing an isothermal surface corresponding to a midplane radius of $\sim$ 40 AU, reproduces the DCO$^+$ emission in the inner disk where the CD channel is not yet active. The ratio of the amount of DCO$^+$ produced by the WD and CD channels  outside the CO snowline in this model is 0.2, consistent with previous full chemical models of DCO$^+$.

\item With the CD channel tracing the CO abundance at 2.0$\times$10$^{-7}$ for radii $<150$ AU, DCO$^+$ is a tracer of the onset of CO evaporation. Full evaporation of CO occurs at radii $<90$ AU, where temperatures are too high for the CD channel to be efficient. This opens possibilities to probe the binding energy of CO ice and its evaporation process.

\item We conclude that the formation and destruction mechanisms of DCO$^+$ are very temperature sensitive, both through the efficiency of the CD channel and the CO abundance. With proper treatment of the DCO$^+$ production through the WD channel, DCO$^+$ can be used as tracer of the location of the CO snowline and the temperature structure, and specifically its gradient, in the outer disk. Furthermore, since the temperature structure and CO abundance both depend on the dust size distribution and spatial distribution, DCO$^+$ is a powerful probe of the dust evolution.
\end{itemize}

\begin{acknowledgements}
The authors acknowledge support
by Allegro, the European ALMA Regional Center node in The Netherlands, and
expert advice from Luke Maud in particular. We also thank Prof. Karin \"Oberg and 
Dr. Stefano Facchini for their very useful discussions that helped improve this paper. This work
was partially supported by grants from the Netherlands Organization for
Scientific Research (NWO) and the Netherlands Research School for Astronomy
(NOVA) This paper makes use of the following
ALMA data: ADS/JAO.ALMA\# 2013.1.01268.S.
ALMA is a partnership of ESO (representing its member states), NSF (USA)
and NINS (Japan), together with NRC (Canada), NSC and ASIAA (Taiwan),
and KASI (Republic of Korea), in cooperation with the Republic of Chile. The
Joint ALMA Observatory is operated by ESO, AUI/NRAO and NAOJ.
\end{acknowledgements}

%
%


\bibliographystyle{aa}     
\bibliography{bibHD163}

\begin{thebibliography}{38}
\expandafter\ifx\csname natexlab\endcsname\relax\def\natexlab#1{#1}\fi

\bibitem[{{Albertsson} {et~al.}(2013){Albertsson}, {Semenov}, {Vasyunin},
  {Henning}, \& {Herbst}}]{Albertsson2013}
{Albertsson}, T., {Semenov}, D.~A., {Vasyunin}, A.~I., {Henning}, T., \&
  {Herbst}, E. 2013, \apjs, 207, 27

\bibitem[{{Brinch} \& {Hogerheijde}(2010)}]{Brinch2010}
{Brinch}, C. \& {Hogerheijde}, M.~R. 2010, \aap, 523, A25

\bibitem[{{Carney} {et~al.}(2017){Carney}, {Hogerheijde}, {Loomis}, {Salinas},
  {{\"O}berg}, {Qi}, \& {Wilner}}]{Carney2017}
{Carney}, M.~T., {Hogerheijde}, M.~R., {Loomis}, R.~A., {et~al.} 2017, \aap,
  605, A21

\bibitem[{{Cleeves}(2016)}]{Cleeves2016}
{Cleeves}, L.~I. 2016, \apjl, 816, L21

\bibitem[{{Dullemond} \& {Dominik}(2004)}]{Dullemond2004b}
{Dullemond}, C.~P. \& {Dominik}, C. 2004, \aap, 417, 159

\bibitem[{{Facchini} {et~al.}(2017){Facchini}, {Birnstiel}, {Bruderer}, \& {van
  Dishoeck}}]{Facchini2017}
{Facchini}, S., {Birnstiel}, T., {Bruderer}, S., \& {van Dishoeck}, E.~F. 2017,
  \aap, 605, A16

\bibitem[{{Favre} {et~al.}(2015){Favre}, {Bergin}, {Cleeves}, {Hersant}, {Qi},
  \& {Aikawa}}]{Favre2015}
{Favre}, C., {Bergin}, E.~A., {Cleeves}, L.~I., {et~al.} 2015, \apjl, 802, L23

\bibitem[{{Flower}(1999)}]{Flower1999}
{Flower}, D.~R. 1999, \mnras, 305, 651

\bibitem[{{Gerner} {et~al.}(2015){Gerner}, {Shirley}, {Beuther}, {Semenov},
  {Linz}, {Albertsson}, \& {Henning}}]{Gerner2015}
{Gerner}, T., {Shirley}, Y.~L., {Beuther}, H., {et~al.} 2015, \aap, 579, A80

\bibitem[{{Guilloteau} {et~al.}(2006){Guilloteau}, {Pi{\'e}tu}, {Dutrey}, \&
  {Gu{\'e}lin}}]{Guilloteau2006}
{Guilloteau}, S., {Pi{\'e}tu}, V., {Dutrey}, A., \& {Gu{\'e}lin}, M. 2006,
  \aap, 448, L5

\bibitem[{{Huang} {et~al.}(2017){Huang}, {{\"O}berg}, {Qi}, {Aikawa},
  {Andrews}, {Furuya}, {Guzm{\'a}n}, {Loomis}, {van Dishoeck}, \&
  {Wilner}}]{Huang2017}
{Huang}, J., {{\"O}berg}, K.~I., {Qi}, C., {et~al.} 2017, \apj, 835, 231

\bibitem[{Isella {et~al.}(2016)Isella, Guidi, Testi, Liu, Li, Li, Weaver,
  Boehler, Carperter, De~Gregorio-Monsalvo, Manara, Natta, P\'erez, Ricci,
  Sargent, Tazzari, \& Turner}]{Isella2016}
Isella, A., Guidi, G., Testi, L., {et~al.} 2016, Phys. Rev. Lett., 117, 251101

\bibitem[{{Kama} {et~al.}(2016){Kama}, {Bruderer}, {van Dishoeck},
  {Hogerheijde}, {Folsom}, {Miotello}, {Fedele}, {Belloche}, {G{\"u}sten}, \&
  {Wyrowski}}]{Kama2016}
{Kama}, M., {Bruderer}, S., {van Dishoeck}, E.~F., {et~al.} 2016, \aap, 592,
  A83

\bibitem[{{Mathews} {et~al.}(2013){Mathews}, {Klaassen}, {Juh{\'a}sz},
  {Harsono}, {Chapillon}, {van Dishoeck}, {Espada}, {de Gregorio-Monsalvo},
  {Hales}, {Hogerheijde}, {Mottram}, {Rawlings}, {Takahashi}, \&
  {Testi}}]{Mathews2013}
{Mathews}, G.~S., {Klaassen}, P.~D., {Juh{\'a}sz}, A., {et~al.} 2013, \aap,
  557, A132

\bibitem[{{Millar} {et~al.}(1989){Millar}, {Bennett}, \& {Herbst}}]{Millar1989}
{Millar}, T.~J., {Bennett}, A., \& {Herbst}, E. 1989, \apj, 340, 906

\bibitem[{{Miotello} {et~al.}(2014){Miotello}, {Bruderer}, \& {van
  Dishoeck}}]{Miotello2014}
{Miotello}, A., {Bruderer}, S., \& {van Dishoeck}, E.~F. 2014, \aap, 572, A96

\bibitem[{{Murillo} {et~al.}(2015){Murillo}, {Bruderer}, {van Dishoeck},
  {Walsh}, {Harsono}, {Lai}, \& {Fuchs}}]{Murillo2015}
{Murillo}, N.~M., {Bruderer}, S., {van Dishoeck}, E.~F., {et~al.} 2015, \aap,
  579, A114

\bibitem[{{{\"O}berg} {et~al.}(2011){{\"O}berg}, {Boogert}, {Pontoppidan}, {van
  den Broek}, {van Dishoeck}, {Bottinelli}, {Blake}, \& {Evans}}]{Oberg2011}
{{\"O}berg}, K.~I., {Boogert}, A.~C.~A., {Pontoppidan}, K.~M., {et~al.} 2011,
  \apj, 740, 109

\bibitem[{{{\"O}berg} {et~al.}(2015){{\"O}berg}, {Furuya}, {Loomis}, {Aikawa},
  {Andrews}, {Qi}, {van Dishoeck}, \& {Wilner}}]{Oberg2015}
{{\"O}berg}, K.~I., {Furuya}, K., {Loomis}, R., {et~al.} 2015, \apj, 810, 112

\bibitem[{{{\"O}berg} {et~al.}(2010){{\"O}berg}, {Qi}, {Fogel}, {Bergin},
  {Andrews}, {Espaillat}, {van Kempen}, {Wilner}, \& {Pascucci}}]{Oberg2010}
{{\"O}berg}, K.~I., {Qi}, C., {Fogel}, J.~K.~J., {et~al.} 2010, \apj, 720, 480

\bibitem[{{Qi} {et~al.}(2011){Qi}, {D'Alessio}, {{\"O}berg}, {Wilner},
  {Hughes}, {Andrews}, \& {Ayala}}]{Qi2011}
{Qi}, C., {D'Alessio}, P., {{\"O}berg}, K.~I., {et~al.} 2011, \apj, 740, 84

\bibitem[{{Qi} {et~al.}(2015){Qi}, {{\"O}berg}, {Andrews}, {Wilner}, {Bergin},
  {Hughes}, {Hogherheijde}, \& {D'Alessio}}]{Qi2015}
{Qi}, C., {{\"O}berg}, K.~I., {Andrews}, S.~M., {et~al.} 2015, \apj, 813, 128

\bibitem[{{Qi} {et~al.}(2013){Qi}, {{\"O}berg}, {Wilner}, {D'Alessio},
  {Bergin}, {Andrews}, {Blake}, {Hogerheijde}, \& {van Dishoeck}}]{Qi2013}
{Qi}, C., {{\"O}berg}, K.~I., {Wilner}, D.~J., {et~al.} 2013, Science, 341, 630

\bibitem[{{Qi} {et~al.}(2008){Qi}, {Wilner}, {Aikawa}, {Blake}, \&
  {Hogerheijde}}]{Qi2008}
{Qi}, C., {Wilner}, D.~J., {Aikawa}, Y., {Blake}, G.~A., \& {Hogerheijde},
  M.~R. 2008, \apj, 681, 1396

\bibitem[{{Remijan} {et~al.}(2003){Remijan}, {Snyder}, {Friedel}, {Liu}, \&
  {Shah}}]{Remijan2003}
{Remijan}, A., {Snyder}, L.~E., {Friedel}, D.~N., {Liu}, S.-Y., \& {Shah},
  R.~Y. 2003, \apj, 590, 314

\bibitem[{{Roueff} {et~al.}(2013){Roueff}, {Gerin}, {Lis}, {Wootten},
  {Marcelino}, {Cernicharo}, \& {Tercero}}]{Roueff2013}
{Roueff}, E., {Gerin}, M., {Lis}, D.~C., {et~al.} 2013, Journal of Physical
  Chemistry A, 117, 9959

\bibitem[{{Salinas} {et~al.}(2017){Salinas}, {Hogerheijde}, {Mathews},
  {{\"O}berg}, {Qi}, {Williams}, \& {Wilner}}]{Salinas2017}
{Salinas}, V.~N., {Hogerheijde}, M.~R., {Mathews}, G.~S., {et~al.} 2017, \aap,
  606, A125

\bibitem[{{Sch{\"o}ier} {et~al.}(2005){Sch{\"o}ier}, {van der Tak}, {van
  Dishoeck}, \& {Black}}]{LAMBDA_database2005}
{Sch{\"o}ier}, F.~L., {van der Tak}, F.~F.~S., {van Dishoeck}, E.~F., \&
  {Black}, J.~H. 2005, \aap, 432, 369

\bibitem[{{Teague} {et~al.}(2015){Teague}, {Semenov}, {Guilloteau}, {Henning},
  {Dutrey}, {Wakelam}, {Chapillon}, \& {Pietu}}]{Teague2015}
{Teague}, R., {Semenov}, D., {Guilloteau}, S., {et~al.} 2015, \aap, 574, A137

\bibitem[{{Turner}(2001)}]{Turner2001}
{Turner}, B.~E. 2001, \apjs, 136, 579

\bibitem[{{van den Ancker} {et~al.}(1998){van den Ancker}, {de Winter}, \&
  {Tjin A Djie}}]{Ancker1998}
{van den Ancker}, M.~E., {de Winter}, D., \& {Tjin A Djie}, H.~R.~E. 1998,
  \aap, 330, 145

\bibitem[{{van Dishoeck} {et~al.}(2003){van Dishoeck}, {Thi}, \& {van
  Zadelhoff}}]{vanDishoeck2003}
{van Dishoeck}, E.~F., {Thi}, W.-F., \& {van Zadelhoff}, G.-J. 2003, \aap, 400,
  L1

\bibitem[{{Vidal-Madjar}(1991)}]{Vidal-Madjar1991}
{Vidal-Madjar}, A. 1991, Advances in Space Research, 11, 97

\bibitem[{{Visser} {et~al.}(2009){Visser}, {van Dishoeck}, \&
  {Black}}]{Visser2009}
{Visser}, R., {van Dishoeck}, E.~F., \& {Black}, J.~H. 2009, \aap, 503, 323

\bibitem[{{Watson}(1976)}]{Watson1976}
{Watson}, W.~D. 1976, Reviews of Modern Physics, 48, 513

\bibitem[{{Willacy}(2007)}]{Willacy2007}
{Willacy}, K. 2007, \apj, 660, 441

\bibitem[{{Wootten}(1987)}]{Wootten1987}
{Wootten}, A. 1987, in IAU Symposium, Vol. 120, Astrochemistry, ed. M.~S.
  {Vardya} \& S.~P. {Tarafdar}, 311--318

\bibitem[{{Yen} {et~al.}(2016){Yen}, {Koch}, {Liu}, {Puspitaningrum}, {Hirano},
  {Lee}, \& {Takakuwa}}]{Yen2016}
{Yen}, H.-W., {Koch}, P.~M., {Liu}, H.~B., {et~al.} 2016, \apj, 832, 204

\end{thebibliography}

\begin{appendix}
\section{Residual channel maps}
\label{appendix}
\begin{figure*}[!ht]
\centering
\begin{subfigure}{0.99\textwidth}
        \centering
        \includegraphics[width=0.95\textwidth]{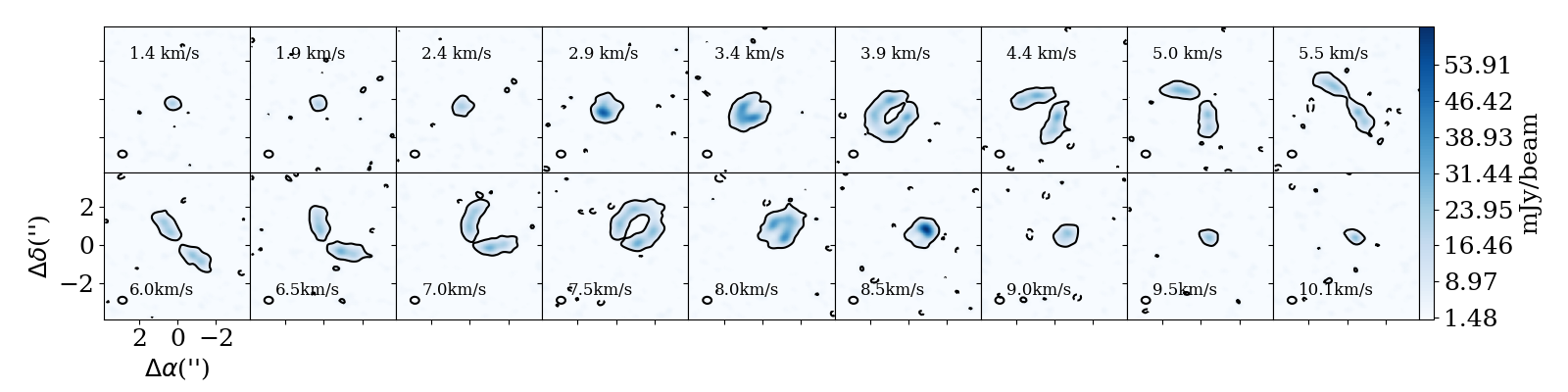}
\end{subfigure}
\begin{subfigure}{0.99\textwidth}
        \centering
        \includegraphics[width=0.95\textwidth]{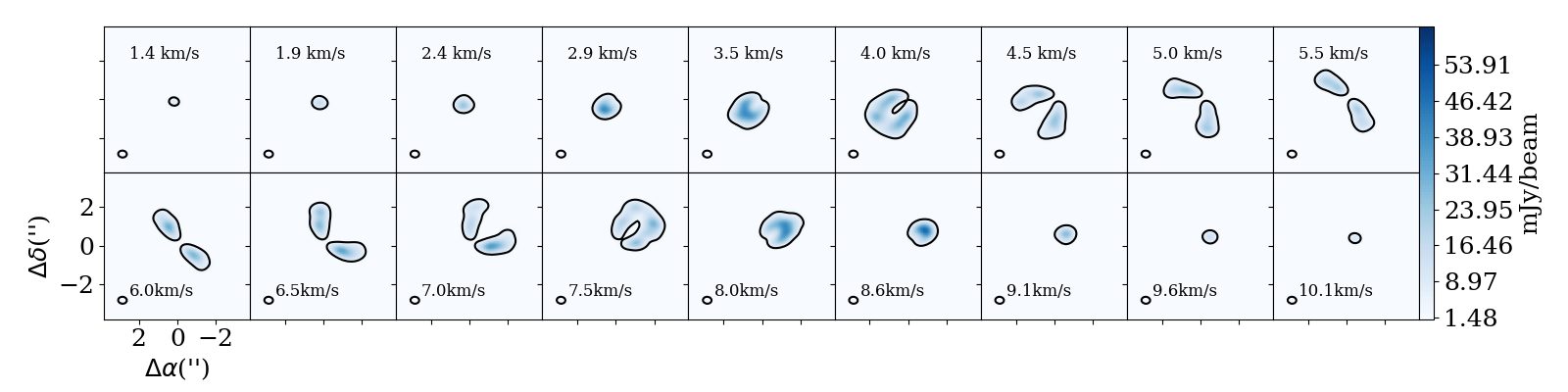}
\end{subfigure}
\begin{subfigure}{0.99\textwidth}		
        \centering
        \includegraphics[width=0.95\textwidth]{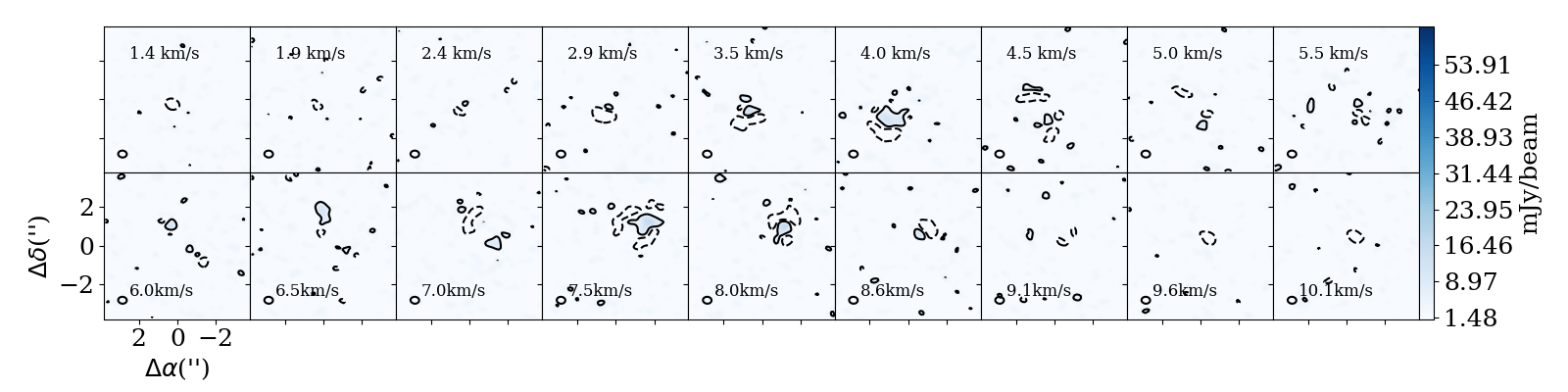}        
\end{subfigure} 
\caption{Top panel shows the DCO$^+$ $J$=3--2 data channel maps. The middle panel map shows the CD+WD+TI model. The bottom panel shows the residual channels. Contours are 3$\sigma$, where $\sigma$ corresponds to the standard deviation of a line free channel. The channel maps presented here are binned in velocity to enhance signal to noise.}
\label{fig:masking}

\end{figure*}
\end{appendix}

\end{document}